\documentclass[floatfix,onecolumn,pre]{revtex4}
\usepackage[pdftex]{graphicx}
\usepackage{amssymb}
\usepackage{amsmath}
\usepackage{latexsym}
\usepackage{bm}
\usepackage{times,subfigure}
\begin{document}

\title{Interfacial motion in flexo- and order-electric switching between nematic filled states}

\author{M. L. Blow}
\author{M. M. Telo da Gama}
\affiliation{Centro de F\'{i}sica Te\'{o}rica e Computacional, Avenida Professor Gama Pinto 2, P-1649-003 Lisboa, Portugal, \\
and \\
Departamento de F\'{i}sica, Faculdade de Ci\^{e}ncias da Universidade de Lisboa, P-1749-016 Lisboa, Portugal}
\date{\today}


\begin{abstract}
We consider a nematic liquid crystal, in coexistence with its isotropic phase, in contact with a substrate patterned with rectangular grooves. In such a system, the nematic phase may fill the grooves without the occurrence of complete wetting. There may exist multiple (meta)stable filled states, each characterised by the type of distortion (bend or splay) in each corner of the groove and by the shape of the nematic-isotropic interface, and additionally the plateaux that separate the grooves may be either dry or wet with a thin layer of nematic. Using numerical simulations, we analyse the dynamical response of the system to an externally-applied electric field, with the aim of identifying switching transitions between these filled states. We find that order-electric coupling between the fluid and the field provides a means of switching between states where the plateaux between grooves are dry and states where they are wet by a nematic layer, without affecting the configuration of the nematic within the groove. We find that flexoelectric coupling may change the nematic texture in the groove, provided that the flexoelectric coupling differentiates between the types of distortion at the corners of the substrate. We identify intermediate stages of the transitions, and the role played by the motion of the nematic-isotropic interface. We determine quantitatively the field magnitudes and orientations required to effect each type of transition.
\end{abstract}
\maketitle



\section{Introduction}
\label{sec:Introduction}

The multiple wetting configurations of nematics on patterned substrates are of great interest in the development of display devices, in particular multistable devices, where two or more distinct display states exist in the absence of applied voltage, allowing greatly enhanced energy efficiency and portability over conventional monostable devices. Multistable devices exploit the multiple textures that result from the interplay between surface geometry and elastic distortion energy of the spatially varying director field.\\

There have been numerous theoretical and simulational studies of field-induced switching of nematics on substrates with periodic patterning. Many have focused on topographically-structured substrates with sinusoidal~\cite{BarberoSkacej,ParryJones,SpencerCare,HarnauDietrich,ParryJonesMeyer} or triangular~\cite{DavidsonBrown} ridges, based on the design of the zenithal bistable nematic device (ZBND)~\cite{BryanBrown}, whilst others have considered flat substrates where the anchoring is spatially varying~\cite{DennistonYeomans,TiribocchiGonnella,AthertonAdler} or inherently bistable~\cite{DavidsonMottram2002}.\\

Structured substrates are also of great current research interest due to their wetting properties. In particular, the control and manipulation of interfaces is utilised in a wide range of microfluidic devices and smart surfaces. In addition to being completely dry or completely wet, a structure may be `filled', with an interface pinned on the topography~\cite{BicoTordeux,IshinoOkumara}. Simple fluids at sinusoidal surfaces~\cite{RasconParry1999} and in wedges~\cite{RasconParry2000} exhibit a filling transition where the interface unbends, followed by a wetting transition where the interface unbinds. Controllable undulation of this interface is being exploited for various applications, for example as a switchable diffraction grating~\cite{BrownWells,BrownMcHale}, or curvature-driven particle assembly~\cite{CavallaroBotto}.\\

When the wetting fluid is nematic, the behaviour is additionally enriched by the long-range interactions between the substrate and the interface, which are mediated by the elasticity of distortions in the nematic's orientation, and by the orientation-dependent effects (anchoring) at the interface itself. On such surfaces, multiple filled states are achievable, with interfaces characterised by differing shape profiles. The interface shape reveals the existence of distinct nematic textures within the grooves. First-order transitions between these states may occur, driven by the anchoring strength or the substrate topography~\cite{PatricioSilvestre,simplyComplicated,SilvestreEskandari}.\\

In this paper, we examine how we may switch between distinct metastable filled states with the use of electric fields. In particular, the role of the interface in the switching dynamics will be investigated. We focus our attention on the coupling of the electric field to gradients in the nematic order. Such coupling is of two main types: coupling to gradients in the rotational orientation, termed flexoelectricity and first identified by Meyer~\cite{Meyer}, and coupling to gradients in the degree of nematic order, termed order-electricity and first identified by Prost and Marcerou~\cite{ProstMarcerou}.\\

We find that order-electricity is a viable mechanism for wetting whilst specific flexoelectric coupling is required to switch between nematic filled states with distinct textures. However, even in this case the dynamics can depend drastically on the presence of the interface, which facilitates the re-orientation of the director field throughout the nematic domain.\\

We model the system using Landau-de-Gennes theory with Beris-Edwards dynamics, numerically solving the equations of motion with the hybrid lattice-Boltzmann/finite-difference method. The paper is arranged as follows. In Sec.~\ref{sec:model} we describe the statics and dynamics of a model nematic in an externally-applied electric field, in Sec.~\ref{sec:system} we characterise the system, and in Sec.~\ref{sec:simulations} we outline the simulation method and parameters. In Sec.~\ref{sec:transitions} we present qualitative and quantitative results for the switching transitions between filled states driven by order-electric and flexoelectric couplings. We first show, in Sec.~\ref{sec:polarisationInterface}, that order-electricity may be used to wet and dewet the substrate. Such switching is akin to electrowetting, which is extensively documented for isotropic fluids~\cite{MugeleBaret}. Then, in Sec.~\ref{sec:polarisationBulk}, we consider how flexoelectricity allows switching between filled states with different topological arrangements of the orientational order. We provide a quantitative analysis of the transitions, identifying the symmetry and magnitude of applied fields that will produce them, and the free-energy barriers and timescales involved. We conclude in Sec.~\ref{sec:discussion} with a discussion and perspectives for future work.

\section{The model}
\label{sec:model}

In order to model phase separation, interfaces and defects, the required nematic order parameter is a traceless, symmetric tensor~\cite{deGennesProst},
\begin{equation}
Q_{\alpha\beta}=\tfrac{1}{2}S\left(3n_{\alpha}n_{\beta}-\delta_{\alpha\beta}\right)+\tfrac{1}{2}B\left(m_{\alpha}m_{\beta}-l_{\alpha}l_{\beta}\right)\;,   \label{eqn:QTensor}
\end{equation}

where $\left\{\mathbf{n},\mathbf{m},\mathbf{l}\right\}$ is an orthonormal set, $S$ is the degree of nematic order along $\mathbf{n}$, and $B$ is the biaxial order along $\mathbf{m}$.\\

The Landau-de-Gennes free energy functional, defined over the fluid domain $\mathcal{D}$ and its boundary with the substrate $\mathcal{\partial D}$, is given by,

\begin{equation}
\mathcal{F}=\int_{\mathcal{D}}\left(f_{\text{bulk}}+f_{\text{elast}}+f_{\text{diel}}+f_{\text{flexo}}\right)d^{3}\mathbf{r}+\int_{\partial\mathcal{D}}f_{\text{wet}}d^{2}\mathbf{r}.    \label{eqn:freeEnergy}
\end{equation}

where 

\begin{align}
f_{\text{bulk}}&=\tfrac{2}{3}A\left\{\tau^{*}S_{\text{nem}}^{-2}Q_{\alpha\beta}Q_{\beta\alpha}-\tfrac{4}{3}(2+\tau^{*})S_{\text{nem}}^{-3}Q_{\alpha\beta}Q_{\beta\gamma}Q_{\gamma\alpha}+\tfrac{2}{3}S_{\text{nem}}^{-4}\left[Q_{\alpha\beta}Q_{\beta\alpha}\right]^{2}\right\},   \label{eqn:freeEnergyBulk} \\
f_{\text{elast}}&=\tfrac{1}{2}L_{1}\partial_{\gamma}Q_{\alpha\beta}\partial_{\gamma}Q_{\alpha\beta}+\tfrac{1}{2}L_{2}\partial_{\alpha}Q_{\alpha\gamma}\partial_{\beta}Q_{\gamma\beta}+\tfrac{1}{2}L_{3}\partial_{\beta}Q_{\alpha\gamma}\partial_{\alpha}Q_{\gamma\beta},    \label{eqn:freeEnergyElas} \\
f_{\text{diel}}&=-\tfrac{1}{2}\epsilon_{0}\left(\epsilon_{\mathrm{I}}\delta_{\alpha\beta} + \epsilon_{\mathrm{A}}Q_{\alpha\beta}\right)E_{\alpha}E_{\beta},   \label{eqn:freeEnergyDielectric} \\
f_{\text{flexo}}&=-\tfrac{4}{3}\chi_{0}E_{\alpha}\partial_{\beta}Q_{\alpha\beta} - \tfrac{4}{3}\chi_{+}E_{\alpha}\partial_{\beta}\left(Q_{\alpha\gamma}Q_{\gamma\beta}\right) - \tfrac{1}{3}\chi_{2}E_{\gamma}\partial_{\gamma}\left(Q_{\alpha\beta}Q_{\beta\alpha}\right) - \tfrac{4}{9}\chi_{-}E_{\alpha}\left(Q_{\alpha\gamma}\partial_{\beta}Q_{\gamma\beta} - Q_{\gamma\beta}\partial_{\beta}Q_{\alpha\gamma}\right),   \label{eqn:freeEnergyFlexo} \\
f_{\text{wet}}&=-\tfrac{2}{3}w_{1}Q_{\alpha\beta}^{\mathrm{S}}Q_{\alpha\beta}+\tfrac{1}{3}w_{2}Q_{\alpha\beta}Q_{\beta\alpha}.  \label{eqn:freeEnergySurface}
\end{align}

$A$ is a positive constant and $\tau^{*}$ is a dimensionless parameter related to temperature. Substitution of  Eqn.~(\ref{eqn:QTensor}) into (\ref{eqn:freeEnergyBulk}) reveals two minima, which correspond to an isotropic phase with $S=S_{\text{iso}}=0$ and a uniaxial nematic phase with $S=S_{\text{nem}}$. The free energy difference between the two phases is
\begin{equation}
f_{\text{bulk}}(S_{\text{nem}})-f_{\text{bulk}}(S_{\text{iso}})=\tfrac{1}{3}A\left(\tau^{*}-1\right)
\end{equation}

Throughout this paper, we shall consider the case of coexistence between isotropic and nematic, $\tau^{*}=1$, but we point out that our qualitative findings should remain applicable slightly above and below coexistence. \\

In the bulk nematic phase, the $L_{i}$ relate to the Frank elastic constants by

\begin{equation}
K_{1}=K_{3}=\tfrac{9}{4}S^{2}_{\text{nem}}\left(2L_{1}+L_{2}+L_{3}\right),\;\;\;\;\;\;\;K_{2}=\tfrac{9}{2}S^{2}_{\text{nem}}L_{1}.  \label{eqn:FrankConstants}  
\end{equation}

In addition to imposing a free energy penalty to gradients in orientation, the elastic terms also penalise gradients in order parameter. At interfaces between the nematic and isotropic phases, the balance of elastic and bulk contributions leads to an interfacial profile with a characteristic width (correlation length) given by

\begin{equation}
\xi=S_{\text{nem}}\sqrt{\frac{3L_{1}+2(L_{2}+L_{3})}{2A}}, \label{eqn:correlationLength}
\end{equation}

and surface tensions,

\begin{equation}
\gamma_{\perp}=\tfrac{1}{6}S_{\text{nem}}\sqrt{A\left(6L_{1}+4(L_{2}+L_{3})\right)},\;\;\;\;\;\;\;\gamma_{\parallel}=\tfrac{1}{6}S_{\text{nem}}\sqrt{A\left(6L_{1}+L_{2}+L_{3}\right)}        \label{eqn:surfaceTension}
\end{equation}

for the cases of homeotropic and planar alignment respectively. Comparing the surface tensions, it is seen that the preference of alignment, or anchoring, is dependent on the sign of $L_{2}+L_{3}$ relative to $L_{1}$.\\

$f_{\text{diel}}$ is the dielectric contribution to the free energy, consisting of the usual isotropic part, with relative permittivity $\epsilon_{\mathrm{I}}$, and a deviatoric part proportional to $\mathbf{Q}$. $f_{\text{flexo}}$ relates to flexoelectricity and order-electricity, and comprises four independent terms when taking gradients of $\mathbf{Q}$ up to quadratic order. It is useful to compare the polarisation described in $f_{\text{flexo}}$ to a phenomenological description in the uniaxial framework~\cite{BarberoDozov},
\begin{equation}
P_{\alpha}=e_{+}S\partial_{\beta}\left(n_{\alpha}n_{\beta}\right)+e_{-}S\left(n_{\alpha}\partial_{\beta}n_{\beta}-n_{\beta}\partial_{\beta}n_{\alpha}\right)+e_{\mathrm{I}}\partial_{\alpha}S+e_{\mathrm{A}}\left(2n_{\alpha}n_{\beta}-\delta_{\alpha\beta}\right)\partial_{\beta}S  \label{eqn:polarisationFrank}
\end{equation}
$e_{\pm}$ are the flexoelectric coefficients, corresponding to polarisations arising from the respective alignment and opposition of splay and bend~\cite{footnoteCoeffs}. $e_{\mathrm{I}}$ and $e_{\mathrm{A}}$ are the order-electric coefficients~\cite{BarberoDozov}. The $e_{\mathrm{I}}$ term purely couples to the gradient of $S$, describing a polarisation across the interface independent of the alignment. The $e_{\mathrm{A}}$ term is alignment dependent, such that homoetropic and planar configurations have opposite polarities. \\

Substitution of Eqn.~(\ref{eqn:QTensor}) into Eqn.(~\ref{eqn:freeEnergyFlexo}), and comparison with Eqn.~(\ref{eqn:polarisationFrank}), yields~\cite{BarberoDozov}
\begin{equation}
\label{eqn:matchCoeffs}
\begin{split}
e_{+}&=2\chi_{0}+S\chi_{+} \\
e_{-}&=S\chi_{-}  \\
e_{\mathrm{I}}&=\tfrac{1}{3}\chi_{0}+\tfrac{5}{3}S\chi_{+}+S\chi_{2}  \\
e_{\mathrm{A}}&=\chi_{0}+S\chi_{+}  \\
\end{split}
\end{equation}
A set of $\{\chi\}$ can be found to satisfy a general choice of $\{e\}$ for most {\em fixed} $S$. But $\{e\}$ cannot be consistently determined when $S$ is varying; an important point of consideration given that $e_{\mathrm{I}}$ and $e_{\mathrm{A}}$ act only on gradients in $S$.  Eqns.~(\ref{eqn:matchCoeffs}) indicate that $\chi_{-}$ and $\chi_{2}$ may be used to independently vary $e_{-}$ and $e_{\mathrm{I}}$ respectively, but it is impossible to set $e_{\mathrm{A}}$ without affecting at least one of $e_{\mathrm{0}}$ and $e_{\mathrm{+}}$. For example, to set a non-zero $e_{\mathrm{A}}$, whilst maintaining $e_{\mathrm{I}}=0$ across the interface, it is necessary to set $\chi_{0}=0$ and $5\chi_{+}=-3\chi_{2}$, in which case $e_{+}$ does not vanish in the bulk. Alternatively, $e_{+}$ may be eliminated in the bulk by setting $\chi_{+}=-2\chi_{0}$, but this does not allow for $e_{\mathrm{I}}$ to be zero everywhere across the interface. To remove these interdependencies, it is necessary to include higher order terms in the polarisation, but this shall not be considered here. The restriction may be likened to the fixed equality between $K_{1}$ and $K_{3}$ expressed in Eqn.~(\ref{eqn:FrankConstants}). \\

Finally, $\mathbf{Q}^{\mathrm{S}}$ represents the approximate order tensor favoured by the substrate, provided that the anchoring strengths $w_{1}$ and $w_{2}$ lie within suitable ranges~\cite{SenSullivan}.\\

Minimisation of the free energy in Eqn.~(\ref{eqn:freeEnergy}), subject to the constraints that $\mathbf{Q}$ be symmetric and traceless, leads to the equilibrium condition $\mathbf{H}^{\text{vol}}=0$ throughout $\mathcal{D}$, and $\mathbf{H}^{\text{surf}}=0$ along $\partial\mathcal{D}$, where

\begin{multline}
H^{\text{vol}}_{\alpha\beta}= \tfrac{4}{3}A\left\{2(2+\tau^{*})S_{\text{nem}}^{-2}\left(Q_{\alpha\gamma}Q_{\gamma\beta}-\tfrac{1}{3}Q_{\gamma\delta}Q_{\delta\gamma}\delta_{\alpha\beta}\right)-\tau^{*}S_{\text{nem}}^{-1}Q_{\alpha\beta}-\tfrac{4}{3}S_{\text{nem}}^{-3}Q_{\gamma\delta}Q_{\delta\gamma}Q_{\alpha\beta}\right\}\\
+ L_{1}\partial_{\gamma\gamma}Q_{\alpha\beta}+(L_{2}+L_{3})\partial_{\gamma\delta}\mathrm{P}_{\alpha\beta\gamma\delta}  \label{eqn:molecularFieldBulk}\\
+\tfrac{1}{2}\epsilon_{0}\epsilon_{\mathrm{A}}(E_{\alpha}E_{\beta}-\tfrac{1}{3}\delta_{\alpha\beta}E^{2}) -\tfrac{4}{3}\chi_{0}\left(\tfrac{1}{2}\partial_{\alpha}E_{\beta}+\tfrac{1}{2}\partial_{\beta}E_{\alpha} -\tfrac{1}{3}\delta_{\alpha\beta}\partial_{\gamma}E_{\gamma}\right)\\
-\tfrac{2}{3}\chi_{2}Q_{\alpha\beta}\partial_{\gamma}E_{\gamma}-\tfrac{4}{3}\chi_{+}\mathrm{P}_{\alpha\beta\gamma\delta}\left(\partial_{\gamma}E_{\delta}+\partial_{\delta}E_{\gamma}\right)-\tfrac{8}{9}\chi_{-}\left(E_{\delta}\partial_{\gamma}-E_{\gamma}\partial_{\delta}\right)\mathrm{P}_{\alpha\beta\gamma\delta}
\end{multline}
 and,
\begin{multline}
H^{\text{surf}}_{\alpha\beta}= -L_{1}\nu_{\gamma}\partial_{\gamma}Q_{\alpha\beta} - \left(L_{2}\nu_{\gamma}\partial_{\delta}+L_{3}\nu_{\delta}\partial_{\gamma}\right)\mathrm{P}_{\alpha\beta\gamma\delta}+\tfrac{2}{3}w_{1}Q_{\alpha\beta}^{\mathrm{S}} +\tfrac{2}{3}\left(\chi_{2}E_{\gamma}\nu_{\gamma}- w_{2}\right)Q_{\alpha\beta}  \label{eqn:molecularFieldSubstrate}  \\ 
+\tfrac{4}{3}\chi_{0}\left(\tfrac{1}{2}E_{\alpha}\nu_{\beta}+\tfrac{1}{2}\nu_{\beta}E_{\alpha}-\tfrac{1}{3}\delta_{\alpha\beta}E_{\gamma}\nu_{\gamma}\right) \\
+\mathrm{P}_{\alpha\beta\gamma\delta}\left(\tfrac{4}{3}\chi_{+}\left\{\nu_{\gamma}E_{\delta}+E_{\gamma}\nu_{\delta}\right\}+\tfrac{4}{9}\chi_{-}\left\{\nu_{\gamma}E_{\delta}-E_{\gamma}\nu_{\delta}\right\}\right)
\end{multline}
where $\mathrm{P}_{\alpha\beta\gamma\delta}=\tfrac{1}{2}\delta_{\alpha\gamma}Q_{\delta\beta}+\tfrac{1}{2}Q_{\alpha\delta}\delta_{\gamma\beta}-\tfrac{1}{3}\delta_{\alpha\beta}Q_{\gamma\delta}$ and $\pmb{\nu}$ is the inward normal to the substrate. \\

To describe the dynamics of the fluid, we must introduce density and velocity fields $\rho$ and $\mathbf{u}$. The evolution of the fluid with time $t$ is described by the continuty, Navier-Stokes, and Beris-Edwards~\cite{BerisEds} equations.
\begin{align}
\partial_{t}\rho+\partial_{\beta}(\rho u_{\beta})&=0      \label{eqn:continuity} \\
\rho\left(\partial_{t}+u_{\beta}\partial_{\beta}\right)u_{\alpha}&= \partial_{\beta}\left[\rho\left\{2\eta\Lambda_{\alpha\beta}-\delta_{\alpha\beta}c_{\mathrm{s}}^{2}\right\}+\left\{\zeta\Sigma_{\alpha\beta\gamma\delta}+ \mathrm{T}_{\alpha\beta\gamma\delta}\right\}H_{\gamma\delta}^{\text{vol}}\right]-H_{\beta\gamma}^{\text{vol}}\partial_{\alpha}Q_{\gamma\beta}      \label{eqn:navierStokes} \\
\left(\partial_{t}+u_{\gamma}\partial_{\gamma}\right)Q_{\alpha\beta}&=-\zeta\Sigma_{\alpha\beta\gamma\delta}\Lambda_{\gamma\delta}-\mathrm{T}_{\alpha\beta\gamma\delta}\Omega_{\gamma\delta}+\Gamma H_{\alpha\beta}^{\text{vol}} \label{eqn:berisEdwards}
\end{align}
where 
\begin{equation}
\begin{split}
\Sigma_{\alpha\beta\gamma\delta}&=\tfrac{4}{3}S_{\text{nem}}^{-1}Q_{\alpha\beta}Q_{\gamma\delta}-\delta_{\alpha\gamma}(Q_{\delta\beta}+\tfrac{1}{2}S_{\text{nem}}\delta_{\delta\beta})-(Q_{\alpha\delta}+\tfrac{1}{2}S_{\text{nem}}\delta_{\alpha\delta})\delta_{\gamma\beta}+\tfrac{2}{3}\delta_{\alpha\beta}(Q_{\gamma\delta}+\tfrac{1}{2}S_{\text{nem}}\delta_{\gamma\delta}),\\
\mathrm{T}_{\alpha\beta\gamma\delta}&=Q_{\alpha\gamma}\delta_{\beta\delta}-\delta_{\alpha\gamma}Q_{\beta\delta},\;\;\;\;\;\Lambda_{\alpha\beta}=\tfrac{1}{2}\left(\partial_{\beta}u_{\alpha}+\partial_{\alpha}u_{\beta}\right),\;\;\;\;\;\Omega_{\alpha\beta}=\tfrac{1}{2}\left(\partial_{\beta}u_{\alpha}-\partial_{\alpha}u_{\beta}\right).
\end{split}
\end{equation}
$\eta$ is the isotropic part of the kinematic viscosity (there are also anisotropic contributions to the viscosity - the Leslie viscosities - arising from the interplay between Eqns.~(\ref{eqn:navierStokes}-\ref{eqn:berisEdwards})~\cite{DennistonOrlandiniPRE}). $c_{\mathrm{s}}=\tfrac{\Delta x}{\sqrt{3} \Delta t}$ is the speed of sound, where $\Delta x$ is the distance corresponding to the seperation of nearest-neighbour lattice nodes, and  $\Delta t$ is the timestep corresponding to the one iteration of the algorithm.  $\zeta$ is a parameter depending on the molecular details of the liquid crystal and $\Gamma$ is the mobility. The system is closed at $\partial\mathcal{D}$ by the conditions of non-slip and thermodynamic equilibrium,

\begin{align}
(\delta_{\alpha\beta}-\nu_{\alpha}\nu_{\beta})u_{\beta}&=0, \label{eqn:nonslip} \\
H_{\alpha\beta}^{\text{surf}}&=0.  \label{eqn:subtrateEquilibrium}
\end{align}

\section{The system}
\label{sec:system}

\begin{figure}
\centering
\includegraphics[width=160mm]{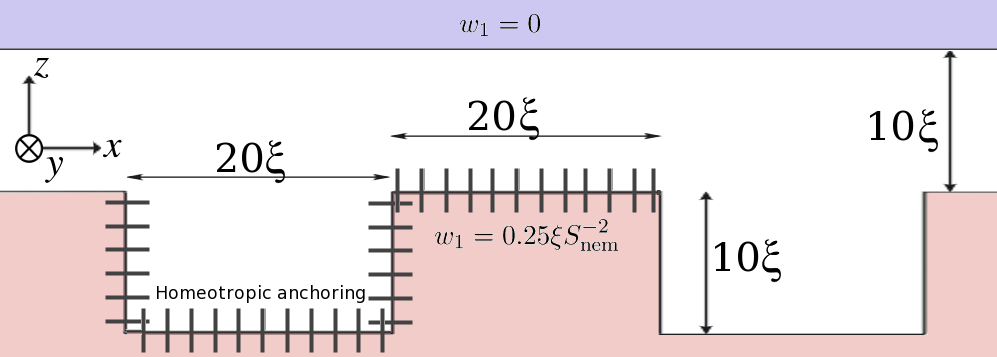}
\caption{Schematic diagram of the substrate studied.}
\label{fig:castellated}
\end{figure}

\begin{figure}
\centering
\includegraphics[width=160mm]{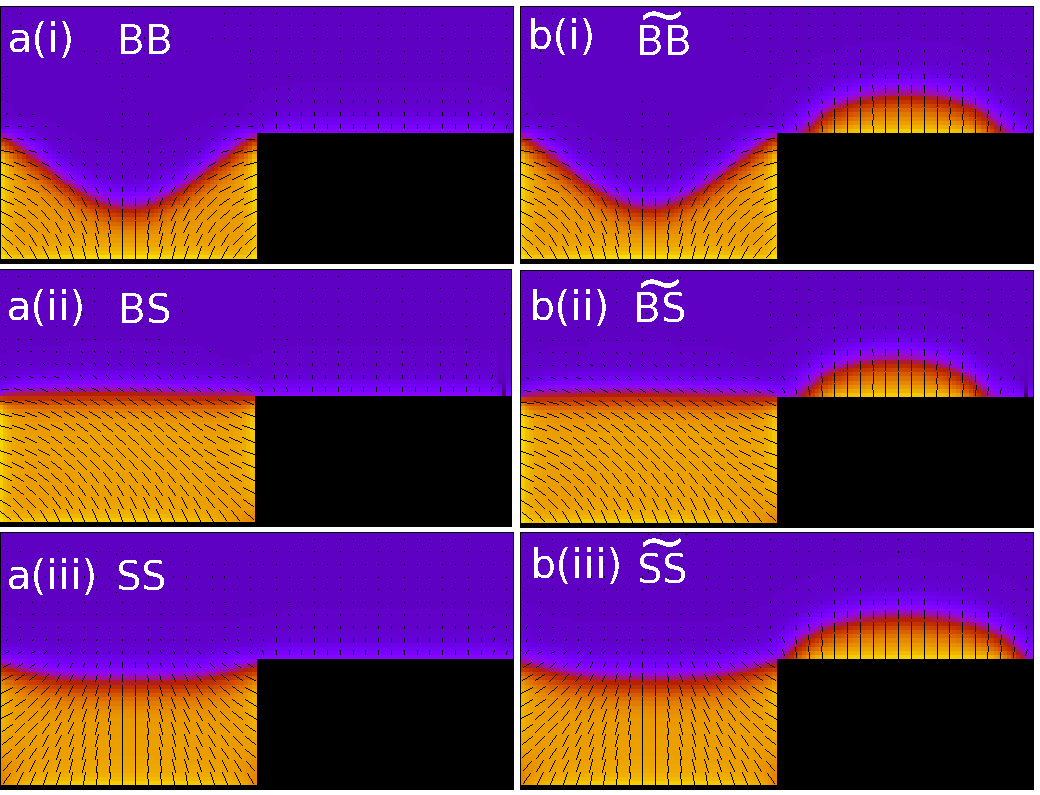}
\caption{(a) States where the nematic phase fills the groove: (i) Bend-Bend, $\mathsf{BB}$, where the interface is bent, (ii) Bend-Splay, $\mathsf{BS}$, where the interface is flat, and (iii) Splay-Splay, $\mathsf{SS}$, where the interface is slightly bent. (b) States where the nematic phase fills the crenel and additionally wets the plateau, but where the interface remains pinned at the upper corners of the groove: (i) $\widetilde{\mathsf{BB}}$, (ii) $\widetilde{\mathsf{BS}}$, (iii) $\widetilde{\mathsf{SS}}$.}
\label{fig:morphologies}
\end{figure}

We consider the system depicted in Fig.~\ref{fig:castellated}. We restrict our study to a two dimensional system confined in the $y$ direction. As such, the system is considered to be two dimensional, lying in the $xz$ plane. In the $x$ direction, the system is periodic, whilst in the $z$ direction, it is confined between two substrates, the lower of which is patterned and has homeotropic anchoring of finite strength. Following~\cite{SilvestreEskandari}, we choose the substrate to be crenellated with rectangular grooves. The grooves have width $20\xi$, and each pair of grooves are separated by a plateau also of width $20\xi$. The height/depth of the grooves is $10\xi$, and the upper substrate is located a further $10\xi$ from the tops of the crenels. This upper substrate is flat and has no anchoring strength ($w_{1}=w_{2}=0$). It exists to close the simulation box in the $+z$ direction, and has no effect on the system energetics or transitions considered in this paper, except for a small set of transitions where the interface touches this upper substrate, nucleating the spreading of the nematic along this substrate.\\

Again following~\cite{SilvestreEskandari}, we choose $L_{2}=2L_{1}$ and $L_{3}=0$. This choice of elastic coefficients gives rise to planar anchoring at the interface. At the substrate, we choose

\begin{equation}
\begin{split}
Q_{\alpha\beta}^{\mathrm{S}}&=\tfrac{3}{2}S_{\text{nem}}\left(\nu_{\alpha}\nu_{\beta}-\tfrac{1}{3}\delta_{\alpha\beta}\right)  \\
w_{1}&=\tfrac{1}{4}\xi S_{\text{nem}}^{-2}   \\
w_{2}&=0
\end{split}
\end{equation}

We focus our attention on states where the base and sides of the grooves are wetted by the nematic phase, but the plateaux and upper surface are in contact with the isotropic fluid. Thus, the interface connects the two upper corners of the groove. We refer to these as `filled' states. Within the groove, the nematic will be distorted due to conflict between the anchoring of the three walls and the interface. At each of the bottom corners, the conflicting anchoring of the rightangled pair of walls causes the director to either bend or splay~\cite{TsakonasDavidson,DavidsonMottram2012,DammoneZacharoudiou}. The combination of these two possibilities in two corners leads to four possible filled states, which we label $\mathsf{BB}$, $\mathsf{SB}$, $\mathsf{BS}$ and $\mathsf{SS}$, where the first and second letters indicate the type of distortion in the left and right lower corners respectively. Of course, $\mathsf{SB}$ and $\mathsf{BS}$ are degenerate, being mirror images.\\

The distortion-mediated interactions between the substrate and nematic-isotropic interface determine the shape of that interface and the relative free energies of the filled states. In the $\mathsf{BB}$ state, the interface follows the contours of the bend, as illustrated in Fig.~\ref{fig:morphologies}a(i), whilst in $\mathsf{SB}$/$\mathsf{BS}$ (Fig.~\ref{fig:morphologies}a(ii)) the interface is flat. In the terminology of~\cite{SilvestreEskandari}, $\mathsf{BB}$ and $\mathsf{SB}$/$\mathsf{BS}$ are thus referred to as the bent and unbent states (referring to the shape of the interface), but $\mathsf{SS}$ is not mentioned, because it is not a global free energy minimum for any of the parameters that the authors consider. Fig.~\ref{fig:morphologies}a(iii) offers an explanation: the configuration in the nematic forces the alignment at the interface to be homeotropic, thereby increasing the surface tension.\\

In these states the plateau between grooves is `dry' (i.e. in contact with the isotropic fluid), but alternatively it may be wetted with a thin layer of nematic, whilst the configuration within the groove is unchanged and the interface remains pinned. Thus there are four further filled states, which we denote $\widetilde{\mathsf{BB}}$, $\widetilde{\mathsf{BS}}$, $\widetilde{\mathsf{SB}}$ and $\widetilde{\mathsf{SS}}$, illustrated in Figs.~\ref{fig:morphologies}b. The interface is pinned at the upper corners of the groove, by virtue of the conflicting orientations either side of the corner.\\

For the parameters specified for this system, \cite{SilvestreEskandari} show that each dry-plateau state has a lower free energy than its corresponding wet-plateau state, and that $\mathsf{BB}$ is the filled state with lowest free energy. We further find that $\mathsf{SB}/\mathsf{BS}$ has a lower free energy than $\mathsf{SS}$.\\

It is possible for the fluid region of the system to be completely isotropic (dry state) or completely nematic. In \cite{SilvestreEskandari}, where an open system is considered, the latter of these includes an interface far from the substrate. In the system considered here, the nematic wets the upper substrate to fill the cell, and is thus confined in a finite space, so the state is not `wet' in the usually understood definition. As with the filled states, there are multiple types of such a state, depending on the nature of the distortion at the four corners. However, since this paper focuses on filled states, where the nematic-isotropic interface is pinned at the upper corners of the grooves.  we presently do not distinguish the `wet' states, and refer to them collectively as $\mathsf{W}$. In Sec.~\ref{sec:discussion}, we shall briefly discuss switching between wet states.\\

\section{Simulation method}
\label{sec:simulations}

To simulate the dynamics of switching, we discretise $\mathcal{D}$ into a square grid of nodes, and evolve the system through a series of time steps, using the hybridised method described in~\cite{HenrichMarenduzzo}. Eqns.~(\ref{eqn:continuity},\ref{eqn:navierStokes}) are evaluated indirectly by a Lattice Boltzmann algorithm - a discretised version of the Boltzmann equation in which the fluid populations are distributed according to the appropriate stresses and propagated between neighbouring nodes~\cite{DennistonOrlandiniPRE,DennistonOrlandiniEPL,HenrichMarenduzzo}.  Eqn.~(\ref{eqn:berisEdwards}) is directly iterated by finite differences. The two algorithms are evaluatated in tandem, with an additional predictor-corrector step to improve stability.\\

Most Lattice Boltzmann studies of liquid crystals have been carried out deep in the nematic phase ($\tau^{*}<1$)\cite{DennistonYeomans,CareHalliday,SpencerCare,HenrichMarenduzzo}. Those dealing with coexisting nematic and isotropic fluids have looked either at two-substance systems, with the relative amounts of each type of material fixed~\cite{SulaimanYeomans}, or single-substance systems at or close to $\tau^{*}=1$~\cite{DennistonOrlandiniEPL}, so that nematic and isotropic coexist in variable proportion. The system we consider is of the latter type.\\

The substrate geometry is a convenient one for the simulation method, with its boundaries lying parallel to the lattice directions. The no-slip condition (\ref{eqn:nonslip}) is enforced using the method of bounce-back on links~\cite{Ladd}, for which it is most amenable to position the boundaries halfway between lattice nodes. Anchoring is implemented differently to previous studies~\cite{SpencerCare,SulaimanYeomans,LintuvuoriMarenduzzo}, requiring the solution of Eqn.~\ref{eqn:subtrateEquilibrium}. This equation is Taylor-expanded in $\mathbf{Q}$. Such a method may be applied to an arbitrarily positioned and orientated substrate.\\

For MBBA, the nematic order parameter at coexistence is $S_{\text{nem}}\approx 0.3$~\cite{SherrellCrellin}, but in the simulations the order is normalised to $S_{\text{nem}}=1$.\\

We use the simulation parameters $A=\tfrac{1}{27}$ and $L_{1}=\tfrac{2}{21}$, which through Eqn.~(\ref{eqn:correlationLength}) produce a correlation length $\xi=3$ lattice spacings, which is found to be sufficient resolution for good agreement in statics with the results of the finite element method employed in \cite{SilvestreEskandari}. Thus the simulation box measures $120$ by $60$ lattice nodes. Viable simulation parameters must be mapped to physical quantities by an appropriate choice of scaling. Using reported data for MBBA~\cite{StinsonLitster1973a,StinsonLitster1973b}, $A\sim 10^{5}\mathrm{Nm^{-2}}$ and $3L_{1}+2L_{2}\sim 10^{-11}\mathrm{N}$, so that the correlation length, $\xi$, and hence the physical lengthscale corresponding to a lattice spacing, is $\Delta x ~\sim10^{-8}\mathrm{m}$. Grooved substrates suitable for liquid crystals with dimensions as small as $5\times10^{-8}\mathrm {m}$ have been produced~\cite{VanDelftVanDenHeuvel,TakahashiSakamoto,LiuLoh}.\\

To ensure stability, the simulation density $\rho$ is limited to a minimum value $\sim 1$. Using the physical value $\rho\sim10^{3}\mathrm{kgm^{-3}}$, we map a simulation mass unit to $10^{-21}\mathrm{kg}$. The timescale may be determined by mapping either surface tension or viscosity. Using Eqn.~(\ref{eqn:surfaceTension}), we derive a value $\gamma=0.022$ in the simulation units and $\gamma\sim10^{-3}\mathrm{Nm^{-1}}$ in physical units. These are matched by one iteration corresponding to $\Delta t\sim 10^{-10}\mathrm{s}$.\\

Using a relaxation time $\tau=2$ in the Lattice Boltzmann algorithm provides an isotropic kinematic viscosity of $0.5$ in lattice units, which corresponds to $\eta\sim 10^{-6}\mathrm{m^{2}s^{-1}}$. This is an order of magnitude smaller than accepted values (generally, stability restrictions make it difficult to provide a compatible viscosity and surface tension, without resorting to a prohibitively large grid~\cite{SulaimanYeomans}). We set $\Gamma=0.45$ and $\zeta=0.4$, providing deviatoric components of viscosity of a like magnitude.\\

Finally, we consider how to incorporate an electric field $\mathbf{E}$ into the simulation. Maxwell's equations state that $\mathbf{E}$ is given by the solution of

\begin{multline}
\epsilon_{0}\left(\epsilon_{\mathrm{I}}\partial_{\alpha}E_{\alpha} + \epsilon_{\mathrm{A}}Q_{\alpha\beta}\partial_{\alpha}E_{\beta}\right)=-\epsilon_{0}\epsilon_{\mathrm{A}}E_{\beta}\partial_{\alpha}Q_{\alpha\beta} + \tfrac{4}{3}\chi_{0}\partial_{\alpha\beta}Q_{\alpha\beta} + \tfrac{4}{3}\chi_{+}\partial_{\alpha\beta}\left(Q_{\alpha\gamma}Q_{\gamma\beta}\right) \label{eqn:Maxwell}  \\
+ \tfrac{1}{3}\chi_{2}\partial_{\alpha\alpha}\left(Q_{\beta\gamma}Q_{\gamma\beta}\right) + \tfrac{4}{9}\chi_{-}\left(\partial_{\alpha}Q_{\alpha\gamma}\partial_{\beta}Q_{\gamma\beta} - \partial_{\alpha}Q_{\gamma\beta}\partial_{\beta}Q_{\alpha\gamma}\right).
\end{multline}

Solution of Eqn.~(\ref{eqn:Maxwell}) at every timestep would add significant computational burden, not least because, unlike iteration of Eqns.~(\ref{eqn:continuity}-\ref{eqn:berisEdwards}), the field cannot be updated using only information from neighbouring nodes. Repeated iterations over the grid would be required at each timestep in order to solve Eqn.~(\ref{eqn:Maxwell}). To avoid this complication, we follow many other numerical studies~\cite{DavidsonBrown,DavidsonMottram2002,DennistonYeomans,SulaimanYeomans,AlexanderYeomans} and experimental analyses~\cite{RyzhkovaPodgornov,LavrentovichLazo} in assuming uniform $\mathbf{E}$. This assumption is consistent with Eqn.~(\ref{eqn:Maxwell}) provided that $\epsilon_{\mathrm{I}}\gg\epsilon_{\mathrm{A}}$ and $\epsilon_{0}\epsilon_{\mathrm{I}}E\gg\chi S_{\text{nem}}\xi^{-1}$.\\

Whether the first of these conditions is met depends solely on the material properties. For example, for MBBA $\epsilon_{\mathrm{A}}/\epsilon_{\mathrm{I}} \sim 0.1$~\cite{deGennesProst}, adequately fulfiling the condition, whilst for 5CB $\epsilon_{\mathrm{A}}/\epsilon_{\mathrm{I}} \sim 10$~\cite{CumminsDunmur}, and thus the condition does not hold.\\

The second condition depends on both the material and the strength of the electric field. Assuming that the first condition is met, the second may be expressed as the ratio

\begin{equation}
r\sim\frac{\xi^{-1}\chi S_{\text{nem}}E}{\epsilon_{0}\epsilon_{\mathrm{I}}E^{2}}\sim\frac{e}{\xi\epsilon_{0}\epsilon_{\mathrm{I}}E}\;.
\end{equation}

We take typical values $\epsilon_{\mathrm{I}}\sim 10$ and $e\sim10^{-11}\mathrm{Cm^{-1}}$~\cite{JewellShambles,CastlesGreen,Ferrarini} in physical units. In Sec.~\ref{sec:transitions}, we shall ascertain that the size of the flexoelectric coupling required to effect transitions is of the order $S_{\text{nem}}eE\sim 10^{-3}\mathrm{Nm^{-1}}$. This requires $E\sim10^{8}\mathrm{Vm^{-1}}$ and results in $r\sim 0.1$, satisfying the second condition. The potential required to produce $E$ on the scale of the device is of the order of volts.\\

The electric field may also be distorted in the vicinity of a non-flat substrate, due to the substrate permittivity being different to $\epsilon_{\mathrm{I}}$. Such an effect is considered in many other works on flexoelectric switching~\cite{HarnauDietrich,ParryJonesMeyer,SpencerCare}, and in some cases plays an essential role in switching of nematic cells. However, it is usually possible to choose materials so that the difference in permittivities is small.

\section{Results}
\label{sec:transitions}

\subsection{Order-electric switching}
\label{sec:polarisationInterface}

\begin{figure}
\centering
\includegraphics[width=160mm]{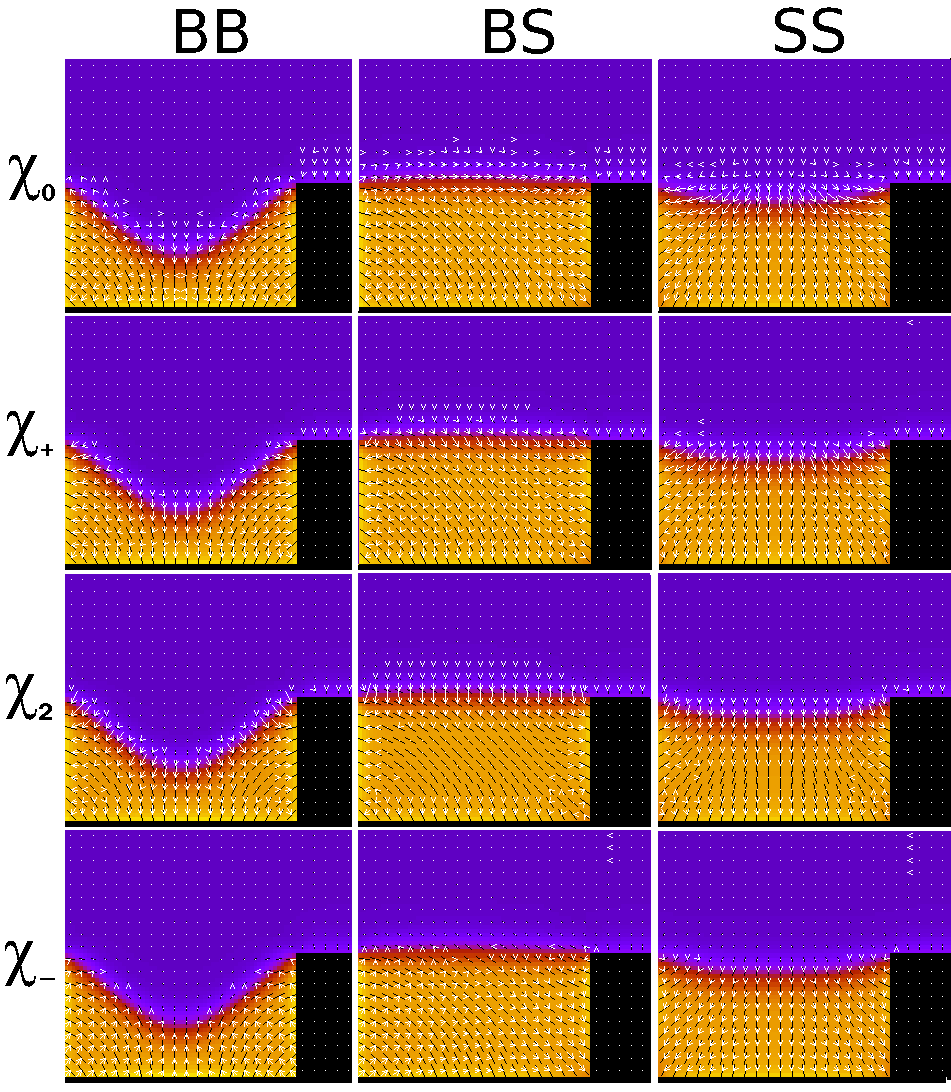}
\caption{The spontaneous polarisations that result in the $\mathsf{BB}$, $\mathsf{BS}$ and $\mathsf{SS}$ states. In each row, the indicated $\chi$ is positive, and the other $\chi$s are all zero.}
\label{fig:flexPolarisations}
\end{figure}

We first consider a system where only $\chi_{2}$ is non-zero, corresponding to a system where there is a polarisation across gradients of $S$, independent of director alignment. The resulting polarisations for the $\mathsf{BB}$ (stable), $\mathsf{BS}$ (metastable), and $\mathsf{SS}$ (metastable), filled states are shown in the third row of Fig.~\ref{fig:flexPolarisations}. \\

\begin{figure}
\centering
\includegraphics[width=160mm]{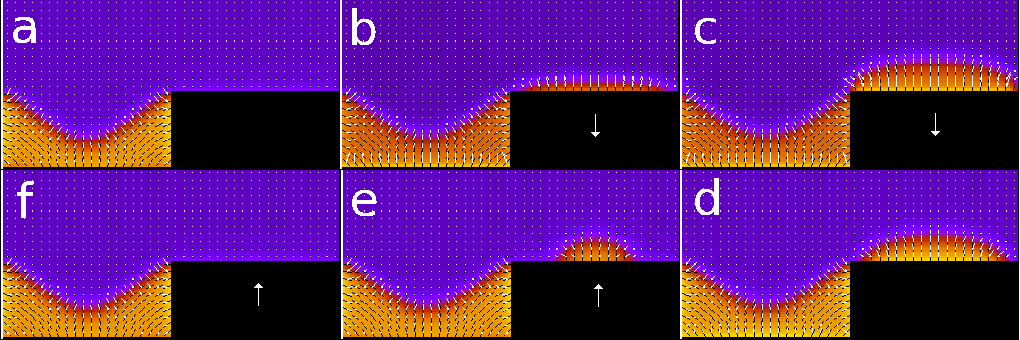}
\caption{(a-d) The transition $\mathsf{BB}\rightarrow\widetilde{\mathsf{BB}}$, through the application of the field $S_{\text{nem}}^{2}\chi_{2}\mathbf{E}=(0,-0.03)$ and (d-f,a) The reverse transition $\widetilde{\mathsf{BB}}\rightarrow\mathsf{BB}$, through $S_{\text{nem}}^{2}\chi_{2}\mathbf{E}=(0,0.03)$. }
\label{fig:switchTop}
\end{figure}

Figs.~\ref{fig:switchTop}a-c show the effect of applying a field $S_{\text{nem}}^{2}\chi_{2}\mathbf{E}=(0,-0.03)$ (normalised by the magnitude of the flexoelectric coefficient and the degree of nematic order) perpendicular to the bottom of the grooves. Starting in the filled state with bend distortion in both lower corners, $\mathsf{BB}$, a homeotropic wetting layer grows on the upper surface of the plateau, until the system resembles $\widetilde{\mathsf{BB}}$. The morphology within the crenel is little changed. When the field is switched off, the system remains in $\widetilde{\mathsf{BB}}$, this being a metastable state at the chosen parameters. Thus, we have effected a transition $\mathsf{BB}\rightarrow\widetilde{\mathsf{BB}}$. \\

The behaviour can be explained as follows. As indicated in Eqn.~(\ref{eqn:molecularFieldSubstrate}), the $\chi_{2}$ gives rise to a surface term with the same form as the quadratic wetting potential, with an effective $w_{2}$ given by $-\chi_{2}\pmb{\nu}.\mathbf{E}$, and thus dependent on the orientation of the substrate relative to the field. On the horizontal sections of the substrate, there is a contribution in $Q_{\alpha\beta}Q_{\beta\alpha}$, which is negative when the substrate faces upwards and hence favours positive $S$, thereby causing the growth of a nematic layer on the plateau. On the vertical sections, however, there is no change to the effective wetting potential, so the contact line does not move. Furthermore, because the field is uniform, there is no force on the polarised interface. \\

By reversing the direction of the field, the sign of the effective $w_{2}$ changes, favouring the isotropic phase $S=0$ at the plateau. Figs.~\ref{fig:switchTop}d-f shows how applying $S_{\text{nem}}^{2}\chi_{2}\mathbf{E}=(0,0.03)$ dewets the plateau. When the field is switched off, the cycle is completed, with the system moving back to the starting configuration of Fig.~\ref{fig:switchTop}a. \\

We observe that transitions between $\mathsf{BS}$ and $\widetilde{\mathsf{BS}}$, and between $\mathsf{SS}$ and $\widetilde{\mathsf{SS}}$ may be brought about by the same means (not shown). \\

We now consider how we might switch the system between the topologies $\mathsf{BB}$, $\mathsf{BS}$ and $\mathsf{SS}$. For example, the transition $\mathsf{BB}\rightarrow\mathsf{BS}$ requires symmetry breaking, indicating that a horizontal field might be used. \\

\begin{figure}
\centering
\includegraphics[width=90mm]{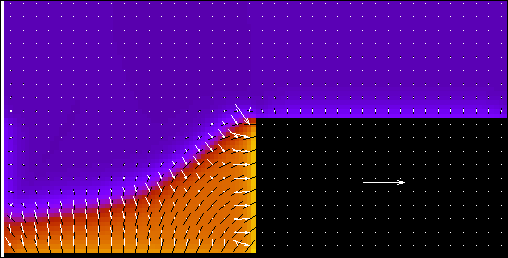}
\caption{The result of applying $S_{\text{nem}}^{2}\chi_{2}\mathbf{E}=(0.03,0)$ to $\mathsf{BB}$}
\label{fig:attemptRightInterface}
\end{figure}

Fig.~\ref{fig:attemptRightInterface} shows the results of applying a field parallel to the base and plateaux surfaces of the grooves $S_{\text{nem}}^{2}\chi_{2}\mathbf{E}=(0.03,0)$. By the same token as above, the lefthand wall of the crenel now favours $S=0$, whilst the righthand wall has (additional) preference for $S=S_{\text{nem}}$. The effect is thus that the lefthand wall dewets, causing the interface to sit diagonally in the crenel. However, when the field is switched off, the nematic rewets the wall, returning to $\mathsf{BB}$ without any transition taking place. \\

Examination of the director orientation in Fig.~\ref{fig:attemptRightInterface} illustrates why this approach will not cause a switch to $\mathsf{BS}$. In each lower corner, the director remains in a bent configuration. It is unsurprising that coupling via $\chi_{2}$ will not bring about a change in the director topology, since it couples only to the degree of ordering $S$.

\subsection{Flexoelectric switching}
\label{sec:polarisationBulk}

\begin{figure}
\centering
\includegraphics[width=160mm]{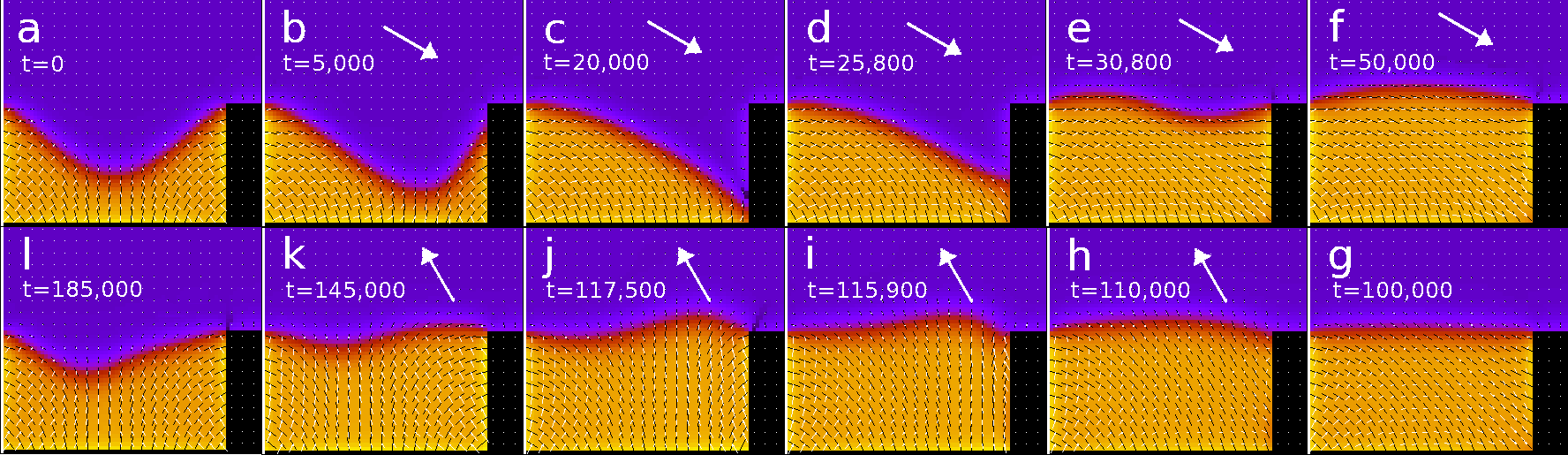}
\caption{(a-h) The transition $\mathsf{BB}\rightarrow\mathsf{BS}$, through the application of the field $S_{\text{nem}}^{2}\chi_{-}\mathbf{E}=(0.0192,-0.0111)$, and (h-l,a) the reverse transition $\mathsf{BS}\rightarrow\mathsf{BB}$, through $S_{\text{nem}}^{2}\chi_{-}\mathbf{E}=(-0.0111,0.0192)$}
\label{fig:switchBB-BS}
\end{figure}

\begin{figure}
\centering
\includegraphics[width=160mm]{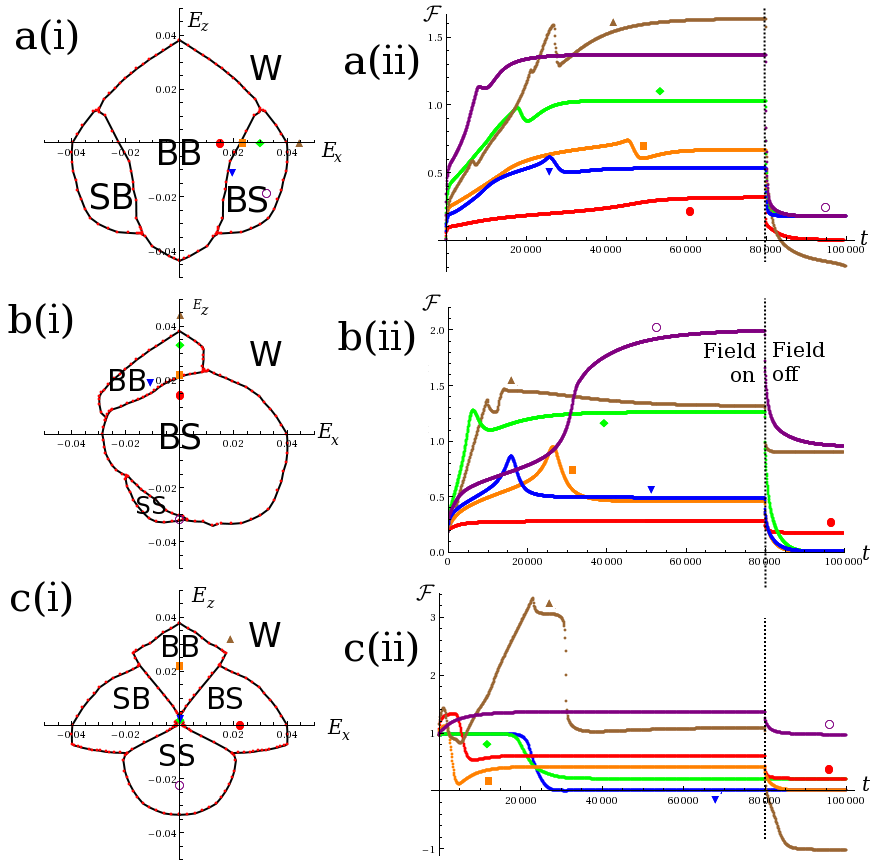}
\caption{(i) The resulting state following the application of an electric field, plotted against $S_{\text{nem}}^{2}\chi_{-}(E_{x},E_{z})$, starting from the state (a) $\widetilde{\mathsf{BB}}$, (b) $\widetilde{\mathsf{BS}}$ and (c) $\widetilde{\mathsf{SS}}$. (ii) Free energy (excluding the electrical contributions) against time, for the fields corresponding to the respective coloured markers in (i). The field is switched on until $t=80,000$ iterations, after which time it is switched off. The blue curves (inverted triangle) in (a) and (b) correspond to the evolution depicted in Fig.~\ref{fig:switchBB-BS}. The purple curve (unfilled circle) in (b) corresponds to Fig.~\ref{fig:switchBS-SS}. The blue (inverted triangle) and green (diamond) curves in (c) represent Figs.~\ref{fig:switchSS-BB} and \ref{fig:switchSS-BS} respectively.}
\label{fig:bulkTransitions}
\end{figure}
\begin{figure}
\centering
\includegraphics[width=160mm]{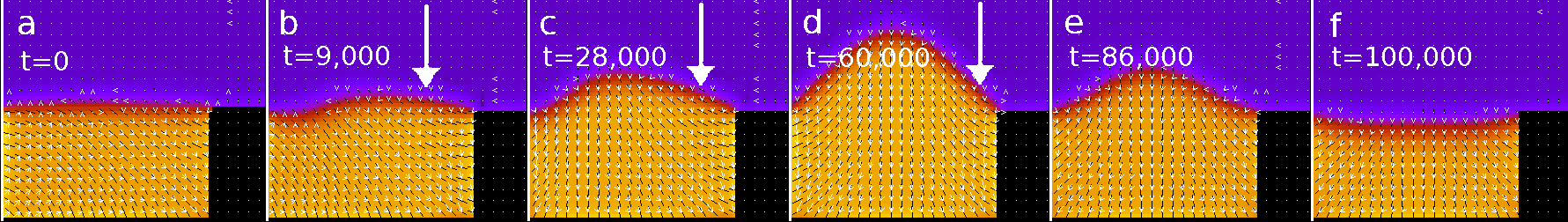}
\caption{The transition $\mathsf{BS}\rightarrow\mathsf{SS}$, through the application of the field $S_{\text{nem}}^{2}\chi_{-}\mathbf{E}=(0,-0.0315)$. The field is switched on in snapshots (a-d) and off in (e-f)}
\label{fig:switchBS-SS}
\end{figure}

We have seen that pure order-electric coupling, through the term with coefficient $\chi_{2}$, is not capable of changing the director topology at the corners of the crenel, and hence bringing about transitions between $\mathsf{BB}$, $\mathsf{BS}$ and $\mathsf{SS}$. We thus turn our attention to the other coefficients present in Eqn.~(\ref{eqn:freeEnergyFlexo}). A comparision is shown in Fig.~\ref{fig:flexPolarisations}.\\

Examining the polarisation in the lower corners of the crenel, we note that for $\chi_{0}$ and $\chi_{+}$ coupling, the polarisation always points outwards at the corners, regardless of whether the director morphology is splayed or bent, with the result that the polarisation fields of the states bear little difference from one another. We therefore conclude that coupling of this nature cannot strongly drive transitions between the states.\\

However, when coupling is of the $\chi_{-}$ form, we see that the polarisation points outwards at corners where the director is bent, but inwards where the director is splayed. As such, the four states $\mathsf{BB}$, $\mathsf{BS}$, $\mathsf{SB}$ and $\mathsf{SS}$ possess different polarisation fields, such that the application of an electric field will favour one over the others. We therefore identify $\chi_{-}$ as the relevent coefficient in driving transitions between the states, and set the other $\chi$ to zero. In terms of the uniaxial coefficients, this corresponds to a system where $e_{+}\ll e_{-}$ in the bulk, as may be found, for example, in certain isomers of MBBA~\cite{Ferrarini}.\\

As shown in Fig.~\ref{fig:flexPolarisations}, in the state $\mathsf{BB}$, the polarisation points outwards from both lower corners, whilst in the state $\mathsf{BS}$, the polarisation in the bottom-left corner also points outwards, but in the bottom-right corner instead points inwards. The application of $\mathbf{E}$ in the downward-right direction would thus be expected to thermodynamically favour $\mathsf{BS}$ over $\mathsf{BB}$, and may offer a means of effecting the transition $\mathsf{BB}\rightarrow\mathsf{BS}$. Figs.~\ref{fig:switchBB-BS}a-f show the effect of applying a field $S_{\text{nem}}^{2}\chi_{-}\mathbf{E}=(0.0192,-0.0111)$, with $\mathsf{BB}$ as the starting configuration shown in Fig.~\ref{fig:switchBB-BS}a. The polarisation in the bottom-right corner opposes the field, so this region of the nematic carries a free energy cost. The interface is thus pulled towards the corner in order to shrink this region, as shown in Fig.~\ref{fig:switchBB-BS}b. This happens until the righthand wall of the crenel has been dewetted, as shown in Fig.~\ref{fig:switchBB-BS}c. At the bottom-right corner, the director rotates anticlockwise, passing through the vertical orientation and continuing until it is again horizontal. The nematic then rewets the wall, as shown in Figs.~\ref{fig:switchBB-BS}d,e, but this time creates a region in which the polarisation is aligned with the field. The system comes to rest in the configuration shown in Fig.~\ref{fig:switchBB-BS}f, which is topologically the same as $\mathsf{BS}$, but with a bulging interface. When the field is swiched off, the system relaxes to $\mathsf{BS}$, as shown in Fig.~\ref{fig:switchBB-BS}g, thus completing the transition $\mathsf{BB}\rightarrow\mathsf{BS}$. \\

In order for the transition to occur, the intensity of the field must be within the correct range. If $\chi_{-}E$ is too small, then the distortion of the interface will not be great enough to allow the director to rotate. For example, switching on the field may produce a state, similar to Fig.~\ref{fig:switchBB-BS}b, that will then relax back to $\mathsf{BB}$ when the field is removed. Alternatively, if $\chi_{-}E$ is too large, then the system will `wet' (i.e. the cell will be completely filled with nematic). This usually occurs by the bulge of the interface in Fig.~\ref{fig:switchBB-BS}f being sufficient for the interface to depin at the upper corners. The distorted nematic has a lower free energy density than the isotropic phase, so wetting of the entire system readily happens once the interface depins.\\

In addition to being a suitable strength, the field must be directed at an appropriate angle. In Fig.~\ref{fig:switchBB-BS}, an angle of $30^{\circ}$ below the horizontal is used, which offers a large range of applicable field strengths, but the transition is possible from $22^{\circ}$ above the horizontal to $68^{\circ}$ below. Fig.~\ref{fig:bulkTransitions}a(i) shows the outcome of applying a field with components $S_{\text{nem}}^{2}\chi_{-}(E_{x},E_{z})$ as marked on the axes. In each case, the field is applied, and the system is left to relax to its new in-field equilibrium. Then the field is removed, and the system is once again left to relax, with the final state being recorded. Of course, the transition $\mathsf{BB}\rightarrow\mathsf{SB}$ occurs under the same criteria as $\mathsf{BB}\rightarrow\mathsf{BS}$, but with the $x$ component of the field reversed. \\

Fig.~\ref{fig:bulkTransitions}a(ii) shows plots of the free energy (excluding the electrical contributions) of the system against time, for the fields corresponding to the respective coloured markers on Fig.~\ref{fig:bulkTransitions}a(i). The E field is kept on until $t=80,000$ iterations, within which the system is fully relaxed in all cases, although in many cases, a much shorter time is sufficient for the transition to proceed. After this time, there is no field. The red curve corresponds to parameters for which the transition does not proceed, and the free energy returns to its original value. The brown curve indicates a situation where the system wets. We note that the wet state actually has the lowest free energy, due to the interface being removed, but it also corresponds to the highest free energy barrier. All other curves correspond to the transition $\mathsf{BB}\rightarrow\mathsf{BS}$. In these cases the final free energy is higher than the original, because $\mathsf{BS}$ is metastable for this system. The blue curve shows the evolution of the system presented in Fig.(\ref{fig:switchBB-BS}). \\

Having outlined the transition $\mathsf{BB}\rightarrow\mathsf{BS}$, we now consider the reverse, $\mathsf{BS}\rightarrow\mathsf{BB}$. As the down-right directed field favoured the polarisation of $\mathsf{BS}$ over $\mathsf{BB}$, an up-left field should favour the reverse. Figs.~\ref{fig:switchBB-BS}g-k show the effect of applying a field $S_{\text{nem}}^{2}\chi_{-}\mathbf{E}=(-0.0111,0.0192)$ to the state $\mathsf{BS}$. Its effect is not to pull the interface, as was the case in the previous transition, but to rotate the director until the geometry is such that the polarisation is aligned to the field. The substrate energy $f_{\text{wet}}$ increases until it reaches a maximum with the director aligned planar to the wall in the vertical orientation (as shown in Fig.~\ref{fig:switchBB-BS}i). Once past this point, $f_{\text{wet}}$ decreases again as the director continues rotating into a new homeotropic alignment, and the system readily evolves until it is in a state topologically similar to $\mathsf{BB}$, as shown in Fig.~\ref{fig:switchBB-BS}j, but with the interface higher and relatively unbent. Switching off the field causes the interface to descend (Fig.~\ref{fig:switchBB-BS}k,l,a), until it has returned to the bent configuration. \\

Fig.~\ref{fig:bulkTransitions}b(i) shows the field response of the system, when initially in state $\mathsf{BS}$. We note that the transition $\mathsf{BS}\rightarrow\mathsf{BB}$ is effected by applying the electric field in the upward and leftward directions, unsurprisingly the opposite of $\mathsf{BB}\rightarrow\mathsf{BS}$, and with a range of fields similar in order of magnitude, if slightly smaller. It is not possible to switch from $\mathsf{BS}$ to its mirror image $\mathsf{SB}$ directly through a single field application. However, it can be done through two successive field applications and relaxations, with $\mathsf{BB}$ as an intermediate state. \\

We notice also a small region in the bottom-left corner of Fig.~\ref{fig:bulkTransitions}b(i), in which the transition $\mathsf{BS}\rightarrow\mathsf{SS}$ occurs. Fig.~\ref{fig:switchBS-SS} outlines the mechanism. Just as an upward field caused the director to rotate on the righthand wall, Figs.~\ref{fig:switchBB-BS}a-c show that a downward field causes a rotation of the director on the lefthand wall, turning the bent texture in the bottom-left corner into splay. The interface then bulges upward, as shown in Fig.~\ref{fig:switchBB-BS}d, and we note that wetting will occur if it reaches the ceiling of the system. Indeed, if the ceiling were higher, then the $\mathsf{BS}\rightarrow\mathsf{SS}$ region of Fig.~\ref{fig:bulkTransitions}b(i) would extend further. When the field is swiched off, this bulge relaxes, as shown in Figs.~\ref{fig:switchBB-BS}e-f.\\

Fig.~\ref{fig:bulkTransitions}b(ii) shows plots of the free energy (excluding the electrical contributions) of the system against time, for the fields corresponding to the respective coloured markers in Fig.~\ref{fig:bulkTransitions}b(i). As with Fig.~\ref{fig:bulkTransitions}a(i), the E field is kept on until $t=80,000$ iterations, and is then switched off. The red curve corresponds to a case where no transition occured, the orange, green and blue curves to various instances of $\mathsf{BS}\rightarrow\mathsf{BB}$, the purple curve to $\mathsf{BS}\rightarrow\mathsf{SS}$, and the brown curve to a scenario where the system wets.\\

Finally, we investigate the response of the state $\mathsf{SS}$ to a field, which is shown in Fig.~\ref{fig:bulkTransitions}c(i). Immediately of note is that, if the field has an upward component, only a very small magnitude is required to enact a transition, either to $\mathsf{BB}$ if the field is close to vertical, or to $\mathsf{BS}$ or $\mathsf{SB}$ if it has a sufficient lateral component. Only against downwardly-orientated fields is $\mathsf{SS}$ robust. Fig.~\ref{fig:bulkTransitions}c(ii) shows the free energy plots. As indicated by the green and blue curves, which correspond to points very close to the origin, there is only a very small energy barrier between $\mathsf{SS}$ and either $\mathsf{BB}$ or $\mathsf{BS}$/$\mathsf{SB}$.

\begin{figure}
\centering
\includegraphics[width=160mm]{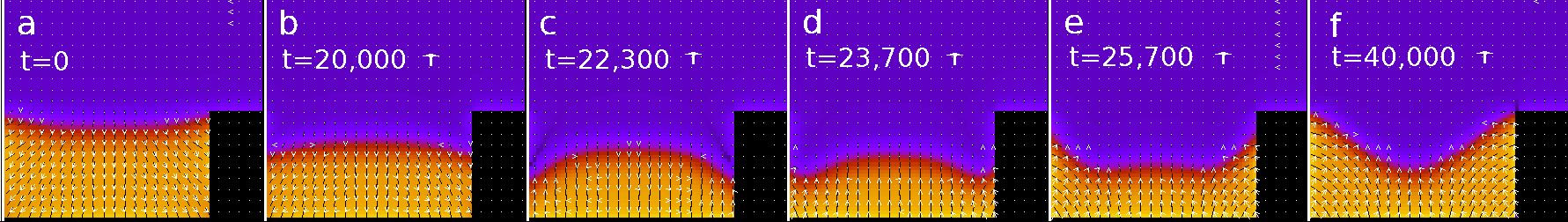}
\caption{The transition $\mathsf{SS}\rightarrow\mathsf{BB}$, through the application of the field $S_{\text{nem}}^{2}\chi_{-}\mathbf{E}=(0,0.00222)$. The field is switched on throughout.}
\label{fig:switchSS-BB}
\end{figure}
\begin{figure}
\centering
\includegraphics[width=160mm]{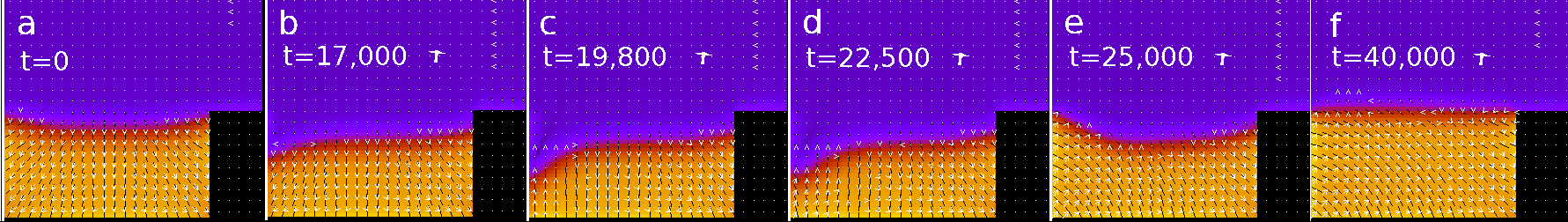}
\caption{The transition $\mathsf{SS}\rightarrow\mathsf{BS}$, through the application of the field $S_{\text{nem}}^{2}\chi_{-}\mathbf{E}=(0.000193,0.00221)$. The field is switched on throughout.}
\label{fig:switchSS-BS}
\end{figure}

\section{Discussion}
\label{sec:discussion}
\begin{figure}
\centering
\includegraphics[width=160mm]{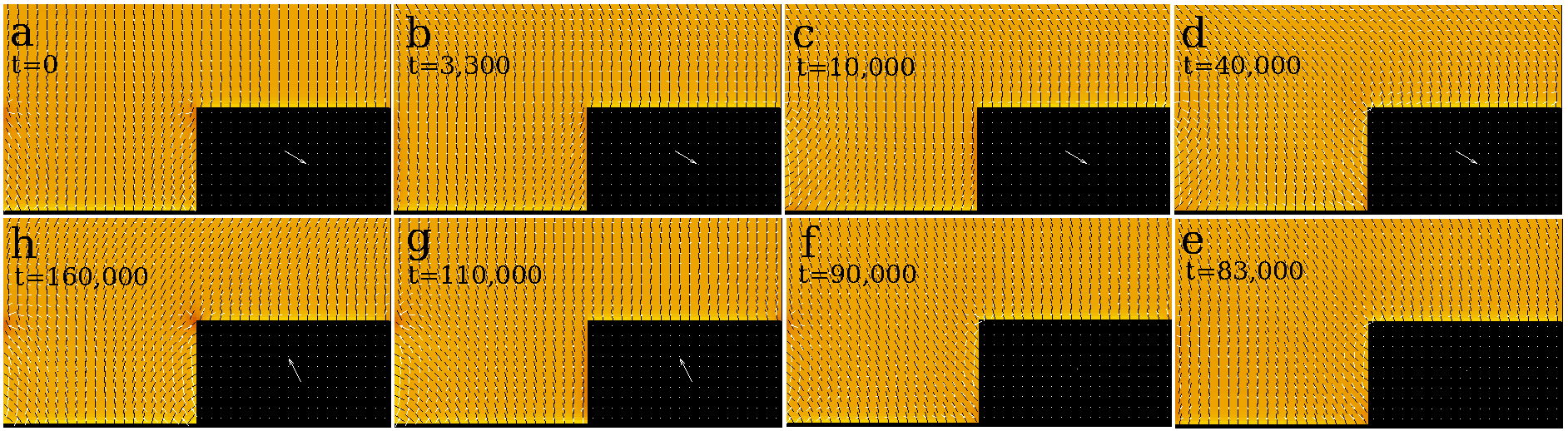}
\caption{Transitions between wet configurations, under electric fields of the same magnitude and directions as applied in Fig.~\ref{fig:switchBB-BS}. (a-f) The transition $\overline{\mathsf{BB}}\rightarrow\overline{\mathsf{BS}}$, through the application of the field $S_{\text{nem}}^{2}\chi_{-}\mathbf{E}=(0.0192,-0.0111)$ (a-d), and subsequent relaxation in the absense of an electric field (d-f). (f-h,a) The reverse transition $\overline{\mathsf{BS}}\rightarrow\overline{\mathsf{BB}}$, through $S_{\text{nem}}^{2}\chi_{-}\mathbf{E}=(-0.0111,0.0192)$.}
\label{fig:wetTransition}
\end{figure}

\begin{figure}
\centering
\includegraphics[width=160mm]{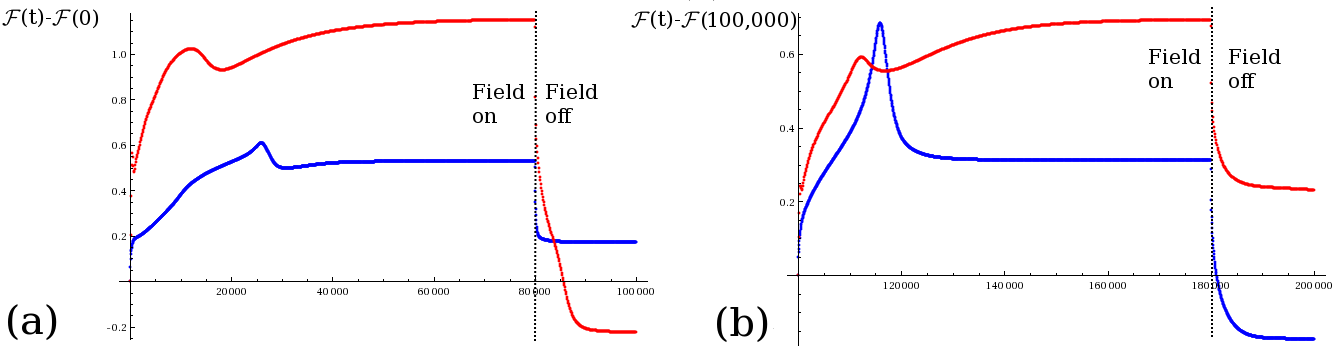}
\caption{Plots of the free energy (excluding electrical contributions) against time for the transitions (blue, a) $\mathsf{BB}\rightarrow\mathsf{BS}$ and (blue, b) $\mathsf{BS}\rightarrow\mathsf{BB}$, depicted in Fig.~\ref{fig:switchBB-BS}, and (red, a) $\overline{\mathsf{BB}}\rightarrow\overline{\mathsf{BS}}$ and (red, b) $\overline{\mathsf{BS}}\rightarrow\overline{\mathsf{BB}}$, shown in Fig.~\ref{fig:wetTransition}. The fields applied are (a) $S_{\text{nem}}^{2}\chi_{-}\mathbf{E}=(0.0192,-0.0111)$, and (b) $S_{\text{nem}}^{2}\chi_{-}\mathbf{E}=(-0.0111,0.0192)$. In order to compare the size of the free energy barriers, the curves are shifted so that both start at the origin in each graph. In fact, the red curve starts with a lower free energy than the blue curve, owing to the absence of interface when the confined geometry is completely filled with nematic.}
\label{fig:comparison}
\end{figure}

Simple fluids at rectangular gratings exhibit a continuous filling transition where the interface unbends, followed by a wetting transition where the interface unbinds~\cite{RasconParry1999}. In nematics, elasticity and topological defects, not present in simple fluids, lead to a variety of novel filled and wet states, characterised by distinct nematic textures~\cite{PatricioSilvestre}.\\

As the geometrical parameters of a rectangular grating vary, the anchoring strength versus roughness nematic surface phase diagram exhibits complex behavior, resulting in the stability of novel filled states not observed in other systems~\cite{SilvestreEskandari}. In particular, two filled states co-exist: the unbent ($\mathsf{SB}$ or $\mathsf{BS}$) filled state, where the NI interface is flat and pinned at the top corners of the grooves, and the bent ($\mathsf{BB}$) state where the NI interface, inside the grooves, is bent. When the system changes from a dry state to the bent filled state, a filling transition occurs but there is no unbending transition.\\

There are important open questions concerning the dynamics of these transitions. The wetting and filling transitions at rectangular gratings are first order and understanding the transition dynamics, involving the nucleation of defects at the corners of the grooves and NI interfacial motion is an important open question.\\

In addition, external fields may be used to drive transitions between filled and wet states and provide a convenient means to investigate switching mechanisms among the nematic wet and filled surface states. In this paper we have reported the results of a hybrid lattice-Boltzmann/finite-element approach to simulate field induced transitions between nematic filled states in rectangular gratings. In particular, we investigated how the application of electric fields, with flexoelectric or order-electric coupling to the nematic, may be used to switch between distinct textures. We have found that order-electric coupling enables switching between states where the plateaux of the rectangular grating are dry and those where they are wet, with the nematic texture in the grooves remaining unchanged, whilst flexoelectric coupling, specifically the component with coefficient $\chi_{-}$, allows for switching between the various nematic textures that may fill the groove.\\

Clearly the presence of the interface is of central importance to the transitions induced by order-electricity, but even in the case of flexoelectric coupling, the interface also affects the dynamics. To demonstrate this, we briefly consider transitions between wet states analogous to the filled states, which we denote $\overline{\mathsf{BB}}$, $\overline{\mathsf{BS}}$, $\overline{\mathsf{SB}}$, and $\overline{\mathsf{SS}}$. Fig.~\ref{fig:wetTransition} shows the transition $\overline{\mathsf{BB}}\rightarrow\overline{\mathsf{BS}}$, followed by the reverse transition $\overline{\mathsf{BS}}\rightarrow\overline{\mathsf{BB}}$. The electric fields employed are the same in strength and direction as those used in Fig.~\ref{fig:switchBB-BS}. Fig.~\ref{fig:comparison} compares the free energy barriers encountered for the filled and wet cases.\\

Comparison of Figs.~\ref{fig:switchBB-BS}(a-g) and \ref{fig:wetTransition}(a-f) reveals that the pathway of the transition $\overline{\mathsf{BB}}\rightarrow\overline{\mathsf{BS}}$ is very different to that of $\mathsf{BB}\rightarrow\mathsf{BS}$, passing through the $\overline{\mathsf{SB}}$ and $\overline{\mathsf{SS}}$ textures before relaxing to $\overline{\mathsf{BS}}$. Fig.~\ref{fig:comparison} shows the free energy barrier to be roughly twice as large in the wet case as in the filled case. By contrast, the pathways of $\overline{\mathsf{BS}}\rightarrow\overline{\mathsf{BB}}$ and $\mathsf{BS}\rightarrow\mathsf{BB}$ are similar and the free energy barriers have roughly the same maximum. We conclude that for $\mathsf{BB}\rightarrow\mathsf{BS}$ the motion of the interface aids the transition, while for $\mathsf{BS}\rightarrow\mathsf{BB}$, the interface does not move significantly and thus the transition is similar in nature to $\overline{\mathsf{BS}}\rightarrow\overline{\mathsf{BB}}$.\\

Further work remains to be done on the effect of deviations of the electric field from uniformity, including flexo/order-electric self-polarisation, dielectric anisotropy and differing permittivities of the substrate and fluid. However, we have shown that, given a suitable liquid crystal, these effects are second-order and do not affect the field symmetry driving the transitions.\\

Another avenue for future work is to extend the study to three dimensions. The application of an electric field may induce gradients in $\mathbf{Q}$ along $y$, breaking the translational symmetry to produce periodic variations in the longitudinal direction of the grooves. A difficulty in simulating such a system is that the repeat length along $y$ is imposed by the length of the simulation boxin this direction, and may not match that naturally adopted by the system.\\

In concusion, the results presented here describe a simple realisation of flexoelectric switching between nematic filled states, in restricted (two-dimensional) rectangular gratings. They highlight the role played by the interface in the dynamics of the transitions between distinct nematic filled states.

\section*{Acknowledgements}

We thank D. G. A. L. Aarts, N. R. Bernardino, N. M. Silvestre, J. M. Yeomans, and I. Zacharoudiou for fruitful discussions. We acknowledge the support of the Portuguese Foundation for Science and Technology (FCT) through the grants SFRH/BPD/73028/2010 (MLB), and PEst-OE/FIS/UI0618/2011 and PTDC/FIS/098254/2008 (MLB and MMTdG).


\begin{thebibliography}{99}
\bibitem{footnoteCoeffs}$e_{\pm}=\tfrac{1}{2}(e_{1}\pm e_{3})$ in the more usual notation~\cite{Meyer,deGennesProst}, where $e_{1}$ and $e_{3}$ are the coefficients corresponding to splay and bend distortions respectively.
\end{thebibliography}
\end{document}